\documentclass[a4paper,12pt]{article}
\usepackage[margin=2cm]{geometry}
\usepackage{comment}
\usepackage{amssymb,extarrows,graphicx,subfigure,setspace}
\usepackage{cite}
\usepackage{slashed}
\usepackage{tensor}
\usepackage[toc,page]{appendix}
\usepackage{color}
\usepackage{physics}
\usepackage{hyperref}
\usepackage{dirtytalk}
\hypersetup{colorlinks=true, linkcolor=blue, citecolor=red, linktoc=page}
\makeatother

\usepackage{amsmath}
\newmuskip\pFqmuskip

\newcommand*\pFq[6][8]{%
  \begingroup 
  \pFqmuskip=#1mu\relax
  \mathcode`\,=\string"8000
  \begingroup\lccode`\~=`\,
  \lowercase{\endgroup\let~}\pFqcomma
  {}_{#2}F_{#3}{\left[\genfrac..{0pt}{}{#4}{#5};#6\right]}%
  \endgroup
}
\newcommand{\pFqcomma}{\mskip\pFqmuskip}

\usepackage{mathrsfs}  
\usepackage{hyperref}

\newcommand{\be}{\begin{equation}}
\newcommand{\bea}{\begin{eqnarray}}
\newcommand{\eea}{\end{eqnarray}}
\newcommand{\ba}{\begin{array}}
\newcommand{\ea}{\end{array}}
\newcommand{\ee}{\end{equation}}
\newcommand{\bes}{\begin{equation*}}
\newcommand{\beas}{\begin{eqnarray*}}
\newcommand{\eeas}{\end{eqnarray*}}
\newcommand{\bas}{\begin{array*}}
\newcommand{\eas}{\end{array*}}
\newcommand{\ees}{\end{equation*}}

\numberwithin{equation}{section}
\begin{document}

	\onehalfspacing
	\noindent
	
	\begin{titlepage}
		
		
		\begin{center}
			
\vspace*{10mm}
{\Large {\bf Traversable Wormholes in Constant Curvature \\[0.2cm] Black Holes
}}

			\vspace*{15mm}
			\vspace*{1mm}
 
  {\bf \large   Ankit Anand\footnote{E-mail : Ankitanandp94@gmail.com}$^{\dag}$, Ruben Campos Delgado\footnote{E-mail : ruben.camposdelgado@gmail.com}$^{\ddag}$ and Daris Samart\footnote{E-mail :  darisa@kku.ac.th }$^{\star}$}
  
  \vspace{0.5cm}

{\it $^{\dag}$ Physics Division, School of Basic and Applied Sciences, Galgotias University, Greater Noida 203201, India \\
$^\ddag$ Institut für Theoretische Physik, Leibniz Universität Hannover, Appelstraße 2, 30167 Hannover, Germany \\ 
$^{\star}$Khon Kaen Particle Physics and Cosmology Theory Group (KKPaCT), Department of Physics, Faculty of Science, Khon Kaen University, 123 Mitraphap Rd., Khon Kaen, 40002, Thailand }

\vspace{0.5cm}

\vspace{0.5cm}

\end{center}
		
\begin{abstract}

This paper investigates the massive gauge field within spacetime context from a $\mathbb{Z}_2$ quotient of the constant curvature black hole. We investigate how the matter field's back reaction affects the spacetime geometry, considering perturbations in the metric up to the first order. The stress-energy tensor's expectation value can be precisely calculated by evaluating its pull-back onto the covering space. By appropriately selecting boundary conditions for the massive vector field along a non-contractible cycle of the quotient manifold, achieving a negative average energy along a null geodesic becomes feasible, enabling a traversable wormhole.

\hspace{5 cm}\\
\hspace{5 cm}\\
\hspace{5 cm}\\
\hspace{5 cm}\\
\hspace{5 cm}\\
\hspace{5 cm}\\
\hspace{5 cm}\\
\hspace{5 cm}


\end{abstract}
\end{titlepage}


\newpage

\newpage
\section{Introduction}\label{Sec: Introduction}

Traversable wormholes are a fascinating topic in theoretical physics, and they are widely Barred in the general theory of relativity. This constraint indicates that communications cannot travel via a wormhole at speeds greater than those feasible through normal space. Notably, this contains the extended Schwarzschild solution, which has a wormhole structure. When we view gravity as a dual representation of a quantum system with AdS-like boundary conditions, the AdS-Schwarzschild wormhole appears as an intriguing analogy to the quantum system's thermofield double state. This state is included in the Hilbert space created by the two original quantum system copies. These two quantum systems can be coupled in this context without any constraints. According to a noteworthy discovery by Gao, Jafferis, and Wall\cite{Gao:2016bin}, simple couplings between these two sides can make the wormhole traversable. These phenomena may be understood as a protocol for information transfer between quantum systems, allowing for a novel approach to the gap between quantum mechanics and gravity theory. This wormhole protocol can be viewed as a very basic implementation of the Hayden-Preskill situation \cite{Hayden:2007cs} of information transferring mechanism in quantum systems, in which a significant prior share of entanglement and a few seemingly insignificant bits—in this case, the two-sided couplings—are sufficient to transfer information from one system to another. The preceding entanglement is crucial to the geometrical configuration because it creates the wormhole that the information travels through. The fact that this phenomenon is enabling us to investigate the black hole or wormholes inside is another intriguing aspect. This is because the two systems' interaction causes a disturbance that essentially pushes the horizon back, revealing more of the interior to the outer world.

\quad The BTZ black hole is the solution to the Einstein field equations in a three-dimensional spacetime with a negative cosmological constant. A well-known method for constructing this black hole involves identifying points along the path of a Killing vector within a three-dimensional anti-de Sitter (AdS) space. The BTZ black hole has a topology of $\mathbb{R}^2 \times S^1$, where $\mathbb{R}^2$ denotes a conformal Minkowski space in two dimensions. Following the same way as done in three dimensions, one can construct analogues of the BTZ solution, the so-called constant curvature (CC) black holes in higher ($D \geq 4$) dimensional AdS spaces \cite{Banados:1997df, Holst:1997tm, Banados:1998dc}. In contrast, these black holes exhibit a topology of $\mathbb{R}^{D-1} \times S^1$ in $D$ dimensions, which differs significantly from the standard topology of $\mathbb{R}^2 \times S^{D-2}$ observed in conventional black holes in $D$ dimensions. Moreover, the exterior region of these constant curvatures (CC) black holes is time-dependent, meaning there is no global time-like Killing vector. This absence complicates the analysis of Hawking radiation and thermodynamics related to these black holes. 

\quad Conversely, these spacetimes provide intriguing examples of smooth, time-dependent solutions. They serve as consistent background spacetimes for string theory, at least to leading order, as they are also vacuum solutions to the Einstein field equations with a negative cosmological constant. Furthermore, these spacetimes are time-dependent, with a time-dependent boundary metric, and are asymptotically AdS. This allows for exploring dual strong coupling field theories in time-dependent backgrounds via the AdS/CFT correspondence \cite{Maldacena:1997re, Gubser:1998bc, Witten:1998qj}. In particular, the D-dimensional constant curvature (CC) black holes have a boundary topology of $dS^{D-2} \times S^1$, where $dS^{D-2}$ represents a $(D-2)$-dimensional de Sitter space. According to the AdS/CFT correspondence, these CC black holes act as gravity duals to strong coupling conformal field theories defined on $dS^{D-2} \times S^1$. Finally, as noted in \cite{Cai:2002mr}, these CC black holes strongly connect to the so-called bubbles of nothing in AdS space \cite{Birmingham:2002st, Balasubramanian:2002am}. These bubbles of nothing were generated by analytically continuing black hole solutions such as Schwarzschild, Reissner-Nordström, and Kerr in AdS spaces. Earlier studies, including \cite{Holst:1997tm, Louko:2004bj, Creighton:1997jp}, explored black hole solutions, entropy, and conserved quantities in these spacetimes. \cite{Cai:2002mr} examined the five-dimensional constant curvature black hole, and \cite{Banados:1998dc} analyzed four-dimensional variants in relation to thermal AdS space and supergravity.

\quad Wormholes, described through solutions of Einstein's equations \cite{Einstein:1935tc, Morris:1988tu}, employ a throat to connect distinct spacetimes or widely separated regions within a single spacetime. It has been revealed that within the General Relativity (GR) framework, traversable wormholes could be conceivable if one allows for exotic matter characterized by negative energy density. While this notion is typically implausible, it does not reside inherently beyond the realm of possibility. In the classical sense, wormholes cannot be traversed, meaning that a causal curve cannot pass through the throat. This prohibition extends to the extended Schwarzschild solution, which incorporates a wormhole. However, when we investigate gravity with Anti-de Sitter (AdS)-like boundary conditions as a gravitational duality of a quantum system, the AdS-Schwarzschild wormhole can be interpreted as the dual representation of the thermofield double state within the quantum system. This state resides in the Hilbert space of two identical copies of the original system, and no restrictions prevent the coupling of these two quantum systems. The null energy condition (NEC) typically prevents traversable wormholes, but quantum fluctuations can violate this, necessitating ANEC violation for traversability. GR theories, like those by Ellis and Bronnikov, require exotic matter to achieve this, as supported by the ER=EPR conjecture \cite{Maldacena:2013xja}.

\quad We choose the backgrounds \(M \), and we use its $\mathbb{Z}_2$ quotient to get our covering spacetime as $\Tilde{M}$. $\Tilde{M}$ features black holes with Killing horizons and well-defined Hartle-Hawking states for any quantum fields. This framework includes several situations, including typical wormhole topologies with a fundamental group \(\pi_1 = \mathbb{Z} \), as well as more unusual configurations, which we can call torsion wormholes, as shown by \(\pi_1 = \mathbb{Z}_2 \). The selection of periodic or anti-periodic boundary conditions imposed by this \(\mathbb{Z}_2 \) symmetry influences the nature of the back-reaction in linear quantum fields, which is critical for deciding whether a wormhole becomes traversable. For scalar fields, this distinction depends on whether the isometry \(J \) maps \(\phi(x) \) to \(\phi(Jx) \) or \(-\phi(Jx) \), hence dictating whether \(\phi \) adheres to periodic or anti-periodic boundary conditions on the quotient spacetime \(M \). Either of the choices results in violation of ANEC and results in the wormhole's traversability.

\quad In this paper, we demonstrated that the choice of boundary condition either for a scalar field or gauge fields is not arbitrary, as it directly influences the violation of the Averaged Null Energy Condition (ANEC). We established that one particular choice leads to the possibility of traversable wormholes. The paper is structured as follows. In Section \ref{Sec:Constant Curvature Black Hole}, we delve into the CC black hole. The subsection discusses the CCBH in the $4D$ case.  Moving forward, Section \ref{Sec:4D CCBH wormhole} focuses on the Average Null Energy Condition for massive gauge fields in $4D$ case and checks for different values of $\delta$. Section \ref{Sec:5D CCBH wormhole} focuses on the Average Null Energy Condition for massive gauge fields in $5D$ case and again checks for different values of $\delta$. Finally, in Section \ref{Sec:Conclusion}, we provide the paper's concluding remarks. Appendix \ref{Appendix:t vector and parallel propagator} presents the expressions for tangent vectors and parallel propagators and their operations.

\section{Constant Curvature Black Hole} \label{Sec:Constant Curvature Black Hole}

Let's commence with the $D$-dimensional Einstein-Hilbert action:

\begin{equation}\label{EH Action}
I_{\text{EH}} = \frac{1}{16\pi G} \int_{\mathcal{M}} d^D x \, \sqrt{-g} (R - 2\Lambda) \ ,
\end{equation}

where \( g_{\mu\nu} \) denotes the metric, \(\Lambda= -(D-1)(D-2)\) represents the cosmological constant, and \( R \) is the Ricci scalar. Varying the action \eqref{EH Action} with respect to the metric tensor yields the Einstein field equations along with a boundary term:

\begin{equation}
    \delta I_{\text{EH}} = \frac{1}{16\pi G} \int_{\mathcal{M}} d^D x \, \sqrt{-g} \, \left[R^\mu_\nu - \frac{1}{2} R \delta^\mu_\nu + \Lambda \delta^\mu_\nu \right] \, (g^{-1} \delta g)^\nu_\mu + \int_{\mathcal{M}} d^D x \, \partial_\mu \Theta^\mu,
\end{equation}

where $\Theta^\mu$ is the surface term, which can be computed using the Christoffel symbol. To explore a $D$-dimensional AdS space, we consider it as a hypersurface embedded in a $(D+1)$-dimensional spacetime, satisfying the following condition:
\begin{equation} \label{Hyperboloid}
 - \sum_{i=1}^2 T_i^2 + \sum_{i=1}^{D-2} X_i^2 +1 = 0 \ ,
\end{equation}
This hyperboloid is embedded within a flat $D$-dimensional space, and its metric is given by:
\begin{equation*}
ds^2 = -\sum_{i=1}^2 dT_i^2  + \sum_{i=1}^{D-2} dX_i^2 \ .
\end{equation*}
This hypersurface possesses a Killing vector with components \( \xi^\alpha = r_h \left( T_1, 0, X_1, 0, 0, \cdots, 0\right) \), representing a boost in the \( (T_1, X_1) \) plane, with a norm \( \xi^2 = r_h^2(T_1^2 - X_1^2) \). By incorporating \( \xi^2 \) into Eq. \eqref{Hyperboloid}, one delineates a \( (D-1) \)-dimensional hypersurface within the AdS space. The sign of \( \xi^2 \)—positive, negative, or zero—plays a crucial role in forming black holes.

Let us confine our discussion to four dimensions by introducing appropriate coordinates. These coordinates involve dimensionless variables \((y_\alpha, \phi)\) defined as:
\[
T_1 = \frac{r}{r_h} \cosh(r_h \phi), \quad X_1 = \frac{r}{r_h} \sinh(r_h \phi), \quad X_j = \frac{2 y_j}{1-y^2} \ ,
\]
where the parameters are related by the following expressions:
\[
r = r_h \frac{1+y^2}{1-y^2}, \quad y^2 = -y_0^2 + y_2^2 + y_3^2 \ .
\]
Here, the coordinates \(X_j\) correspond to \(T_2\), \(X_2\), and \(X_3\) for \(y_0\), \(y_2\), and \(y_3\) respectively. The ranges for these coordinates are \(-\infty < y_j < \infty\) and \(-\infty < \phi < \infty\), with the restriction \( -1 < y^2 < 1\). The boundary condition, as \(r\) approaches infinity, corresponds to a hyperbolic ball where \(y^2 = 1\). The induced metric in these coordinates can be expressed as:
\begin{equation}\label{metric in y}
    ds^2 = \frac{(r^2+r_h^2)^2}{r_h^2} \left(-dy_0^2 + dy_2^2 + dy_3^2\right) + r^2 d\phi^2 \ .
\end{equation}
The associated Killing vector field is \(\xi = \partial_\phi\), with \(\xi^2 = r^2\). The quotient space can be identified by the condition \(\phi \equiv \phi + 2n\pi\), leading to a spacetime topology of \(\mathbb{R}^3 \times S^1\). By further introducing coordinates on the hyperplane \((y_0, y_2, y_3)\) as follows:
\[
y_0 = \sqrt{\frac{r-r_h}{r+r_h}} \sinh(r_h t), \quad y_2 = \sqrt{\frac{r-r_h}{r+r_h}} \cosh(r_h t) \cos(\theta), \quad y_3 = \sqrt{\frac{r-r_h}{r+r_h}} \cosh(r_h t) \sin(\theta) \ ,
\]
the metric in \eqref{metric in y} transforms into:
\begin{equation}\label{metric in rt}
ds^2 = - (r^2 - r_h^2) dt^2 + \frac{dr^2}{r^2 - r_h^2} + \frac{r^2 - r_h^2}{r_h^2} \cosh^2(r_h t) d\theta^2 + r^2 d\phi^2 \ .
\end{equation}
This coordinate system reveals the time dependence of the solution, similar to the de Sitter spacetimes. In static coordinates, de Sitter space appears static within the cosmological horizon but does not encompass the entire de Sitter space. In contrast, the solution in global coordinates covers the entire space but is time-dependent. It can be easily verified that \(\beta = \pi/r_h\). Interpreting this as the inverse Hawking temperature of the black hole raises concerns, as the surface gravity \(\kappa = r_h\) does not align with the standard black hole thermodynamics relation \(\beta = \frac{2\pi}{\kappa}\). 

The metric \eqref{metric in rt} can also be represented in Kruskal-like coordinates \((U,V,\theta, \phi)\) as:
\begin{equation}\label{Metric u v}
    ds^2 = \frac{1}{(1 + UV)^2} \left(-4 dU dV + (U-V)^2 d\theta^2 + r_h^2 (1-UV)^2 d\phi^2\right) \ . 
\end{equation}
The boundaries of the black hole's horizons are located at \(U = 0\) and \(V = 0\). The condition \(1 + UV = 0\) corresponds to the spacetime boundaries. The \(\mathbb{Z}_2\) quotient \cite{Louko:2004bj} identification can be expressed as:
\[
(U, V, \theta, \phi) \sim (V, U, \theta, \phi + \pi) \ .
\]
The geodesics distance\footnote{Detailed derivation is outlines in Appendix \ref{Appendix:t vector and parallel propagator}} $\mu$ is 
\begin{eqnarray}\label{Geodesics distance}
	 \cosh {\mu} = \frac{ (U+V)(U'+V')- (U-V)(U'-V') \mathcal{C}_\theta + (-1+U'V')(-1+UV)K}{(1+UV)(1+U'V')}  \ ,
\end{eqnarray}
where, $\mathcal{C}_\theta=\cos\left(r_h[\theta-\theta']\right)$ and $K=\cosh\left(r_h[\phi-\phi']\right)$.

At the linear order\cite{Gao:2016bin}, the geodesic equation for a null ray originating from the far past of the right boundary is given by:
\[
V(U) = \frac{-1}{2g_{\alpha\beta}(V=0)}\int_{\infty}^U dU \; h_{kk} \ ,
\]
where \(h_{kk} = k^\alpha k^\beta h_{\alpha\beta}\) represents the first-order perturbation in the metric, \(g_{\alpha\beta} = g_{\alpha\beta} + \delta g_{\alpha\beta} = g_{\alpha\beta} + \epsilon g_{\alpha\beta} + \mathcal{O}(\epsilon^2)\). Moreover, we consider that the background metric in \eqref{Metric u v} maintains a constant \(g_{\alpha\beta}\) along the horizon (\(V = 0\)). The linearized form of Einstein's equation governing the metric perturbation on the chosen horizon can be expressed as:
\begin{equation}\label{Linearized EE}
    \left[\frac{5}{2} h_{UU} + \frac{3}{2} \partial_U (U h_{UU}) - \frac{1}{2r_h^2} \partial_U^2 h_{\phi\phi}\right] = 16\pi G_N T_{UU} \ .
\end{equation}
Integrating \eqref{Linearized EE} over all \(U\) and neglecting boundary terms to ensure consistency of the boundary stress tensor, the change \(\Delta V\) as \(U\) approaches infinity is given by:
\[
\Delta V(+\infty) \propto \int_{-\infty}^{\infty} dU \langle T_{kk} \rangle \ ,
\]
indicating that for the wormhole to be traversable, one must have \(\Delta V(+\infty) < 0\).


\section{Massive Gauge field and Stress-Energy tensor}

The action for massive gauge field is 
\begin{equation} \label{Massive action}
    \mathcal{S}= \int d^4x \sqrt{-g} \left(-\frac{1}{4} g^{\alpha \beta} g^{\tau \nu} F_{\alpha \tau}F_{\beta \nu} -\frac{1}{2} m^2  g^{\tau \nu} A_\tau A_\nu \right) \ ,
\end{equation}
where $F_{\tau \nu}=\nabla_\tau A_\nu - \nabla_\nu A_\tau$. The equation of motion(EOM) and stress-energy tensor corresponding to action \eqref{Massive action} respectively can be written as
\begin{equation}\label{EOM massive}
     g_{\tau \nu} \left(\Box- \kappa \right) A^\tau = 0 \ , 
     \end{equation}
     where $\Box$ is laplacian, $\kappa=m^2-3$ and 
     \begin{equation}\label{stress-energy tensor}
      T_{\alpha \beta} = g^{\gamma \delta} F_{\gamma \alpha} F_{\delta \beta} + m^2 A_\alpha A_\beta - g_{\alpha \beta} \left[\frac{1}{4}g^{\gamma \delta}g^{\sigma \rho} F_{\gamma \sigma} F_{ \delta \rho}+\frac{1}{2}m^2g^{\gamma \delta} A_\gamma A_\delta \right] \ .
\end{equation}
The term in parenthesis does not contribute by taking the double null component. We will not carry over this term from now on.

The massive vector field related to $\mathcal{M}$ and its  $\mathbb{Z}_2$ quotient spacetime $\Tilde{\mathcal{M}}$ is
\begin{equation}\label{State relation}
    A_\alpha^\pm(x) = \frac{1}{\sqrt{2}}\left\{\Tilde{A}_\alpha(\Tilde{x})\pm \Tilde{A}_{\alpha '}(\Tilde{x}')\right\}  \ ,
\end{equation}
here $\pm$ corresponds to the periodic and anti-periodic boundary conditions. Now, we will use these states to find the double-null component of the stress-energy tensor. The expectation value of the double null component of stress-energy tensor as already shown in \cite{Anand:2020wlk} in Hartle-Hawking-like states(HHS) after pulling back them in covering space is
\begin{eqnarray}\label{Talphabeta}
   _\pm\bra{HHS_\mathcal{M}}k^\alpha k^\beta T_{\alpha \beta}(x)\ket{HHS_\mathcal{M}}_\pm &=& \pm \Bigg[\bra{HHS_{\Tilde{\mathcal{M}}}}g^{\gamma \delta'}(\Tilde{x},\Tilde{x}') k^\mu \Tilde{F}_{\gamma \mu}(\Tilde{x})k^{\beta'} \Tilde{F}_{\delta' \beta'}(\Tilde{x}')\ket{HHS_{\Tilde{\mathcal{M}}}} \nonumber \\
   &&+m^2 \bra{HHS_{\Tilde{\mathcal{M}}}} k^\alpha\Tilde{A}_\alpha(\Tilde{x})k^{\beta'}\Tilde{A}_{\beta'}(\Tilde{x}')\ket{HHS_{\Tilde{\mathcal{M}}}} \Bigg] \ .
\end{eqnarray}
From this, it is easy to see that one of the boundary conditions\footnote{ Here, the boundary conditions refer to the choice of sign in Eq.\eqref{State relation} i.e., either positive or negative choice.} violates the ANEC and results in a traversable wormhole. From \eqref{Talphabeta}, it becomes clear to check the violation of the ANEC validation, and it is necessary to calculate the $\bra{HHS_{\Tilde{\mathcal{M}}}} \Tilde{A}_{\alpha}(\Tilde{x})\Tilde{A}_{\beta'}(\Tilde{x}')\ket{HHS_{\Tilde{\mathcal{M}}}}$ and $\bra{HHS_{\Tilde{\mathcal{M}}}}\Tilde{F}_{\gamma \mu}(\Tilde{x}) \Tilde{F}_{\delta' \beta'}(\Tilde{x}')\ket{HHS_{\Tilde{\mathcal{M}}}}$ within the covering space $\Tilde{\mathcal{M}}$ and to compute this we have first to derive the expression for the two-point function.

Using the co-ordinate discussed in Appendix\ref{Appendix:t vector and parallel propagator}, the expression for the geodesics distance is 
\begin{eqnarray}
  \cosh {\mu} &=& T_1 T_1' + T_2 T_2' - X_1 X_1' - X_2 X_2' - X_3 X_3' 
  \cr && \cr
  &=& \frac{1}{(1+UV)(1+U'V')} \Big[ (U+V)(U'+V')- (U-V)(U'-V') \; \cos \; \left(r_h(\theta-\theta')\right) \nonumber \\
	&& \;\;\;\;\;\;\;\;\;\;\;\;\;\;\;\;\;\;\;\;\;\;\;\;\;\;\;\;\;\;\;\;\;\;\;\;\;\;\;\;\;\;\;\;\;\;\;\;\;\; + (-1+U'V')(-1+UV)\; \cosh \; \left(r_h(\phi-\phi')\right) \Big] \nonumber \ ,
\end{eqnarray} 
where the unprimed and primed coordinates are $(U,V,\theta,\phi)$ and $(U',V',\theta',\phi')$ respectively.


\subsection{The Two-point function}
In this subsection, we briefly go through the derivation of two-point functions in a massive vector field case. The two-point function $Q_{\mu\nu '}(x, x') = \expval{ A_{\mu }(x) A_{\nu '}(x')}$ will also satisfy the same EOM i.e., \eqref{EOM massive} and extra condition as $A^\nu$ satisfies, i.e., $\nabla_\nu A^\nu=0$. For a maximally symmetric state, one can start with the general expression of a two-point function in terms of the $t$-vectors and parallel propagator\footnote{The formulae we will use to calculate the t-vector and parallel propagator is \begin{equation*}
    t_\alpha(x,x')=\nabla_\alpha  \mu\;\;\;;\;\;\;t_{\alpha '}(x,x')=\nabla_{\alpha '}\mu\;\;\;;\;\;\;g_{\alpha\beta '}(x, x')t^{\beta '}(x, x') + t_\alpha(x, x') = 0 \ .
\end{equation*}} as 
\begin{equation}\label{Two point f}
    Q_{\mu\nu '}(x, x') = \mathcal{C}(\mu) g_{\mu \nu'} +\mathcal{D}(\mu) t_\mu t_{\nu'} \ ,
\end{equation}
where $\mathcal{C}(\mu)$ and $\mathcal{D}(\mu)$\footnote{ Following \cite{Allen:1985wd}, we can write the derivatives of fundamental objects as 
 \begin{eqnarray}\label{VW relation}
          \nabla_\alpha t_\beta &=& \mathcal{V}(\mu) \left[g_{\alpha\beta}-t_\alpha t_\beta\right]\;\;\;\;\;;\;\;\;\;\; \nabla_\alpha t_{\beta'} = \mathcal{W}(\mu) \left[g_{\alpha\beta'}+t_\alpha t_{\beta'}\right] \ .
      \end{eqnarray}}
are two unknown functions. We must calculate these two different functions to get the expression for the two-point function. Since the two-point function also satisfies the EOM, by putting \eqref{Two point f} into \eqref{EOM massive}, we have  
\begin{eqnarray}\label{condition11}
0 &=&\frac{d^2 \; \mathcal{C(\mu)}}{d \mu^2}+3 \mathcal{V}(\mu) \frac{d \mathcal{C}(\mu)}{d \mu}-\{[\mathcal{V(\mu)}+\mathcal{W}(\mu)]^2+\kappa\}\mathcal{C}(\mu) + 2 \mathcal{V}(\mu) \mathcal{W}(\mu) \mathcal{D}(\mu)  
\cr && \cr 
0 &=&\frac{d^2\; \mathcal{D}(\mu)}{d \mu^2}+ 3 \mathcal{V}(\mu) \frac{d\;\mathcal{D}(\mu)}{d \mu}+\{2 \mathcal{V}(\mu) \mathcal{W}(\mu) - 3 (\mathcal{V}(\mu)^2+\mathcal{W}(\mu)^2)-\kappa\}\mathcal{D}(\mu) \nonumber \\ \label{condition2}
&&\;\;\;\;\;\;\;\;\;\;\;\;\;\;\;\;\;\;\;\;\;\;\;\;\;\;\;\;\;\;\;\;\;\;\;\;\;\;\;\;\;\;\;\;\;\;\;\;\;\;\;\;\;\;+2(\mathcal{V}(\mu)+\mathcal{W}(\mu))^2\mathcal{C}(\mu) \ ,
\end{eqnarray}
where $\mathcal{V}(\mu)$ and $\mathcal{W}(\mu)$ are already defined in \eqref{VW relation}  and their values in $AdS$ space are $\mathcal{V}(\mu)=\coth {\mu}$ and $\mathcal{W}(\mu)=-\csch {\mu}$. By defining $ {^{4D}\mathcal{L}(\mu)}=\mathcal{C}(\mu)-\mathcal{D}(\mu)$, the above equation reduced to 
 \begin{equation}\label{Extra condition 2}
     \frac{d^2\;{^{4D}\mathcal{L}(\mu)}}{d\mu^2}+3 \coth(\mu) \frac{d\; {^{4D}\mathcal{L}(\mu)}}{d\mu} - \{3+6\csch^2(\mu)+\kappa\} {^{4D}\mathcal{L}(\mu)} - 6 \coth{(\mu)} \csch(\mu) \mathcal{C}(\mu)=0 \ .
 \end{equation}
Finally, the extra condition i.e., $\nabla_\nu Q^{\nu \tau'}$, we have  
\begin{eqnarray}\label{Extra condition}
     \frac{d\; {^{4D}\mathcal{L}(\mu)}}{d\mu}+3 \coth(\mu)  {^{4D}\mathcal{L}(\mu)} - 3\csch(\mu) \mathcal{C}(\mu) = 0 \ .
\end{eqnarray}
 Now, putting \eqref{Extra condition} into \eqref{Extra condition 2} to eliminate $\mathcal{C}(\mu)$, and with the change in variable as $y=\cosh^2(\mu/2)$, then the solution to the above equation\footnote{The Hypergeometric function is
 \begin{equation}\label{Hypergeometric Equation}
    \left\{ y(1-y)\frac{d^2}{dy^2} + \left[\Delta_0-\left(\Delta_+ +\Delta_-+1\right)y\right]\frac{d}{dy}-\Delta_+\Delta_- \right\}  {^{4D}\mathcal{L}(y)}=0 \ .
 \end{equation}} is 
\begin{equation}\label{L(y)}
     {^{4D}\mathcal{L}(y)} = \frac{-3\; \Gamma(\Delta_+) \Gamma(\Delta_+-\Delta_0+1)}{32 \pi^2 \Gamma(\Delta_+-\Delta_-+1)}\frac{1}{m^2}\frac{1}{y^{\Delta_+}} {_2F_1} \left[\Delta_+,\Delta_+- \Delta_0 +1;\Delta_+-\Delta_-+1;\frac{1}{y}\right] \ ,
\end{equation}
with
 \begin{equation}\label{Delta Expression}
     \Delta_\pm = \frac{5\pm \sqrt{1+4m^2}}{2} =  \frac{5\pm \delta}{2}  \;\;\;\;\;\;\;\;\;\;\text{and}\;\;\;\;\;\;\;\;\;\;\Delta_0=3
 \end{equation}
 The expressions of $\mathcal{C}(\mu)$ and $\mathcal{D}(\mu)$ in the terms of $ {^{4D}\mathcal{L}(\mu)}$ are 
 \begin{eqnarray}
     \mathcal{C}(\mu) &=& \frac{1}{3}\sinh{(\mu)} \mathcal{L}'(\mu)+\cosh{(\mu)} {^{4D}\mathcal{L}(\mu)} \ , \nonumber \\ \label{C and D}
     \mathcal{D}(\mu) &=& \frac{1}{3}\sinh{(\mu)} \mathcal{L}'(\mu)+\{\cosh{(\mu)}-1\} {^{4D}\mathcal{L}(\mu)} \ . \nonumber
 \end{eqnarray}

 Now, we aim to compute the field strength, i.e.,  $\expval{F_{\mu \nu}F_{\alpha' \beta'}}$. Let's briefly sketch the proof for the calculation of the second one
\begin{equation}
    \begin{split}
        \expval{F_{\mu \nu}F_{\alpha' \beta'}} &= \expval{(\nabla_\mu A_\nu - \nabla_\nu A_\mu)(\nabla_{\alpha'} A_{\beta'} - \nabla_{\beta'} A_{\alpha'})} = \expval{\nabla_\mu A_\nu \nabla_{\alpha'}A_{\beta'}} + \cdots \\[1ex]
        &= \nabla_\mu \nabla_{\alpha'} \left[\mathcal{C}(\mu) g_{\nu \beta'} +\mathcal{D}(\mu) t_\nu t_{\beta'}\right] + \cdots = \left\{\mathcal{C}''(\mu) + \mathcal{C}'(\mu) \coth(\mu) \right\} t_\mu t_{\alpha'} g_{\nu \beta'} + \cdots \ .  \nonumber
    \end{split}
\end{equation}
     Taking all the terms together, we have 
     \begin{equation}\label{Field Strength}
         \begin{split}
        \expval{F_{\mu \nu}F_{\alpha' \beta'}} &= \mathcal{G_N} \left[g_{\nu \beta'} t_{\mu}  t_{\alpha'} -g_{\mu \beta'} t_{\nu}  t_{\alpha'} + g_{\mu \alpha'} t_{\nu}  t_{\beta'}-g_{\nu \alpha'} t_{\mu}  t_{\beta'} \right] + \mathcal{G_G} \left[g_{\mu \beta'}g_{\nu \alpha'}-g_{\nu \beta'}g_{\mu \alpha'}\right] \ ,
    \end{split}
     \end{equation}
     here we have defined two different functions $\mathcal{G_N}$ and $\mathcal{G_G}$. In the terms of $\mathcal{C}(\mu)$, $\mathcal{C}(\mu)$ and their derivative w.r.t $\mu$ we have
     \begin{eqnarray}
         \mathcal{G_N} &=&  \tanh \left(\frac{\mu }{2}\right) \left(\mathcal{C}'+\mathcal{C} \tanh \left(\frac{\mu }{2}\right)\right)+\text{csch}(\mu ) \left(\mathcal{C}' (\cosh (\mu )-2)+\mathcal{D}' -2 \mathcal{D} \text{csch}(\mu )\right)+\mathcal{C}'' \nonumber \\
        \mathcal{G_G} &=& 2 \text{csch}^2(\mu ) (\mathcal{D}+\mathcal{C} (\cosh (\mu )-1)+\mathcal{C}' \sinh (\mu )) \ . \nonumber
     \end{eqnarray}
     


\subsection{Wormhole in $4D$ CCBH}\label{Sec:4D CCBH wormhole}

This subsection investigates the traversability of the Wormhole in $4D$ CCBH. From eq.\eqref{Talphabeta}, it is clear that to check the ANEC, our first aim is to compute the two-point function. Eq.\eqref{Two point f}, denotes the general expression for the two-point function and there are two unknown parameters one must compute for the expression of the two-point function. At the same time, eq.\eqref{C and D}, have a relation of the two unknowns in terms of the new function $ {^{4D}\mathcal{L}(y)}$, and its general expression is already defined in eq.\eqref{L(y)}. Let's start with the expression for $ {^{4D}\mathcal{L}(\mu)}$ in terms of $\delta$ as defined in \eqref{Delta Expression} as 
 \begin{equation}\label{K mu expression}
    {^{4D}\mathcal{L}(\mu)} =  -\frac{3 \Gamma \left(\frac{\delta }{2}-\frac{1}{2}\right) \Gamma \left(\frac{\delta }{2}+\frac{5}{2}\right)}{16 \pi ^2 \cosh^{\delta +5} \left(\frac{\mu }{2}\right)\Gamma (\delta +2)} \,_2F_1\left[\frac{\delta +1}{2},\frac{\delta +5}{2};\delta +1;\text{sech}^2\left(\frac{\mu }{2}\right)\right] \ .
\end{equation}
With the help of this expression, one can easily calculate the expression of unknown quantities in the two-point function. Just for maintaining the continuity, let's write the expression for the unknowns i.e., $\mathcal{C}(\mu)$ and $\mathcal{D}(\mu)$ as 
\begin{eqnarray}
    \mathcal{C}(\mu) &=& \frac{\Gamma \left(\frac{\delta -1}{2}\right) \Gamma \left(\frac{\delta +5}{2}\right)  }{32 \pi ^2  \cosh ^{5+\delta}\left(\frac{\mu }{2}\right) \Gamma (\delta +2)}\Big(((\delta -1) \cosh (\mu )-\delta-5) \, _2F_1\left(\frac{\delta +1}{2},\frac{\delta +5}{2};\delta +1;\text{sech}^2\left(\frac{\mu }{2}\right)\right) \nonumber \\
    && \;\;\;\;\;\;\;\;\;\;\;\;\;\;\;\;\;\;\;\;\;\;\;\;\;+(\delta +5) \tanh ^2\left(\frac{\mu }{2}\right) \, _2F_1\left(\frac{\delta +3}{2},\frac{\delta +7}{2};\delta +2;\text{sech}^2\left(\frac{\mu }{2}\right)\right)\Big)\nonumber \ ,\\
     \mathcal{D}(\mu) &=& \frac{\Gamma \left(\frac{\delta -1}{2}\right) \Gamma \left(\frac{\delta +5}{2}\right) \sinh ^2\left(\frac{\mu }{2}\right) }{32 \pi ^2 \cosh ^{\delta+7}\left(\frac{\mu }{2}\right)  \Gamma (\delta +2)}\Big((\delta +5) \, _2F_1\left(\frac{\delta +3}{2},\frac{\delta +7}{2};\delta +2;\text{sech}^2\left(\frac{\mu }{2}\right)\right) \nonumber \\
     && \;\;\;\;\;\;\;\;\;\;\;\;\;\;\;\;\;\;\;\;+(\delta -1) (\cosh (\mu )+1) \, _2F_1\left(\frac{\delta +1}{2},\frac{\delta +5}{2};\delta +1;\text{sech}^2\left(\frac{\mu }{2}\right)\right)\Big) \ .\nonumber
\end{eqnarray}

Now, by using these relation we can compute the two-point function as in eq.\eqref{Two point f} and the field strength as in eq.\eqref{Field Strength}. By computing them together we can write them in the terms of $t_\alpha$ and $g_{\alpha \beta'}$ as 
\begin{equation}\label{FFm2}
   \expval{ g^{\gamma \delta'} F_{\gamma \alpha}(x)F_{\delta' \beta'}(x) + m^2 A_\alpha(x) A_{\beta'}(x')} = \mathscr{T} (\mu,\delta)\; t_\alpha t_{\beta'} + \mathscr{G}(\mu,\delta)\; g_{\alpha \beta'} \ ,
\end{equation}
where,
\begin{eqnarray}
        \mathscr{T}(\mu,\delta) &=& \frac{\Gamma \left(\frac{\delta -1}{2}\right) \Gamma \left(\frac{\delta +5}{2}\right)}{512 \pi ^2} \frac{\sinh^2{\frac{\mu}{2}}}{\cosh^{13+\delta}{\frac{\mu}{2}}} \Bigg[4 \cosh ^6\left(\frac{\mu }{2}\right) \left(\left(\delta ^2-1\right) \cosh (\mu )-\delta  (\delta +16)-31\right)  \nonumber \\
        && \, _2\tilde{F}_1\left(\frac{\delta +1}{2},\frac{\delta +5}{2};\delta +1;\text{sech}^2\left(\frac{\mu }{2}\right)\right)+(\delta +5) \Bigg\{(2 \cosh ^4\left(\frac{\mu }{2}\right) ((\delta +1) (3 \delta +5) \cosh (\mu ) \nonumber \\
        && -\delta  (3 \delta +40)-85) \, _2\tilde{F}_1\left(\frac{\delta +3}{2},\frac{\delta +7}{2};\delta +2;\text{sech}^2\left(\frac{\mu }{2}\right)\right)+(\delta +3) (\delta +7) \nonumber \\
        && \Bigg(\cosh ^2\left(\frac{\mu }{2}\right) ((3 \delta +7) \cosh (\mu )-3 \delta -23) \, _2\tilde{F}_1\left(\frac{\delta +5}{2},\frac{\delta +9}{2};\delta +3;\text{sech}^2\left(\frac{\mu }{2}\right)\right) \nonumber \\
        && +(\delta +5) (\delta +9) \sinh ^2\left(\frac{\mu }{2}\right) \, _2\tilde{F}_1\left(\frac{\delta +7}{2},\frac{\delta +11}{2};\delta +4;\text{sech}^2\left(\frac{\mu }{2}\right)\right)\Bigg)\Bigg\}\Bigg]
        \cr && \cr
        \mathscr{G}(\mu,\delta) &=& \frac{\Gamma \left(\frac{\delta -1}{2}\right) \Gamma \left(\frac{\delta +5}{2}\right) }{64 \pi ^2} \text{sech}^{11+\delta}\left(\frac{\mu }{2}\right)  \Bigg[\cosh ^4\left(\frac{\mu }{2}\right) (\delta  (-\cosh (\mu ))+\delta +\cosh (\mu )+11) \nonumber \\
        && \, _2\tilde{F}_1\left(\frac{\delta +1}{2},\frac{\delta +5}{2};\delta +1;\text{sech}^2\left(\frac{\mu }{2}\right)\right)+\frac{1}{2} (\delta +5) \Bigg\{-2 \cosh ^2\left(\frac{\mu }{2}\right) (\delta  \cosh (\mu )-\delta \nonumber \\
        && +\cosh (\mu )-7) \, _2\tilde{F}_1\left(\frac{\delta +3}{2},\frac{\delta +7}{2};\delta +2;\text{sech}^2\left(\frac{\mu }{2}\right)\right)-(\delta +3) (\delta +7) \sinh ^2\left(\frac{\mu }{2}\right) \nonumber \\
        && \, _2\tilde{F}_1\left(\frac{\delta +5}{2},\frac{\delta +9}{2};\delta +3;\text{sech}^2\left(\frac{\mu }{2}\right)\right)\Bigg\}\Bigg] \ . \nonumber
\end{eqnarray}


A detailed study of $\expval{T_{kk}}$ and \( \int \expval{T_{kk}}dU\) remains to be done. For this task, we rely on numerical methods. These expressions are quite complicated, so we will look at different values of $\delta$ and check whether the ANEC is violated.
\section*{Case I: \(\delta = 0\)}

In this scenario, where the parameter \(\delta\) is fixed at zero, we examine the double null component of the stress-energy tensor. The expression for this component is given by:
\begin{equation}
T_{kk}(U, K) = \langle k^\delta k^{\sigma'} (g^{\eta\rho'} F_{\eta\delta}(x) F_{\rho'\sigma'}(x') + m^2 A_\delta(x)A_{\sigma'}(x')) \rangle
\end{equation}
This simplifies to:
\begin{eqnarray}
T_{kk}(U, K) &=& \frac{U^2 \left(K+U^2\right)}{2 \sqrt{2} \pi ^2 \left(K+2 U^2-1\right)^3 \left(K+2 U^2+1\right)^{3/2}} \Bigg\{\left(4K + 8U^2 - 4\right) \mathcal{K} \left(\frac{2}{2 U^2+K+1}\right) \nonumber\\
&& - \left(7K + 14U^2 + 9\right) \mathcal{E} \left(\frac{2}{2 U^2+K+1}\right)\Bigg\}
\end{eqnarray}
where:
\begin{itemize}
    \item \(\mathcal{K}(k)\) denotes the complete elliptic integral of the first kind, and
    \item \(\mathcal{E}(k)\) represents the complete elliptic integral of the second kind.
\end{itemize}

\begin{figure}[ht]
    \begin{center}
        \includegraphics[scale=0.45]{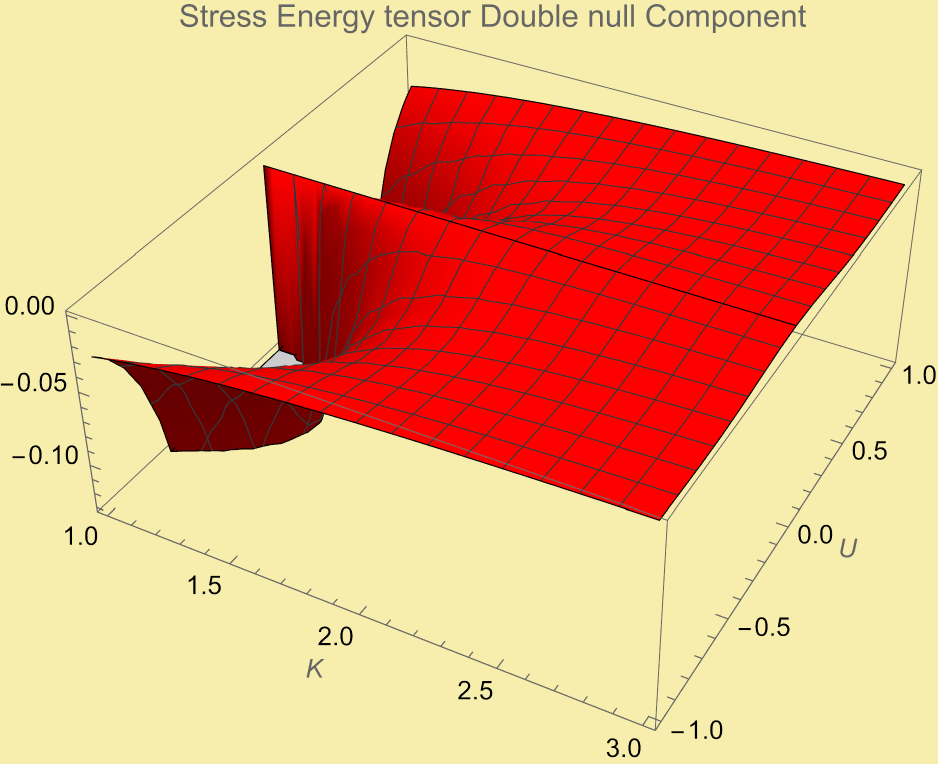}
        \hspace{0.2cm}
        \includegraphics[scale=0.69]{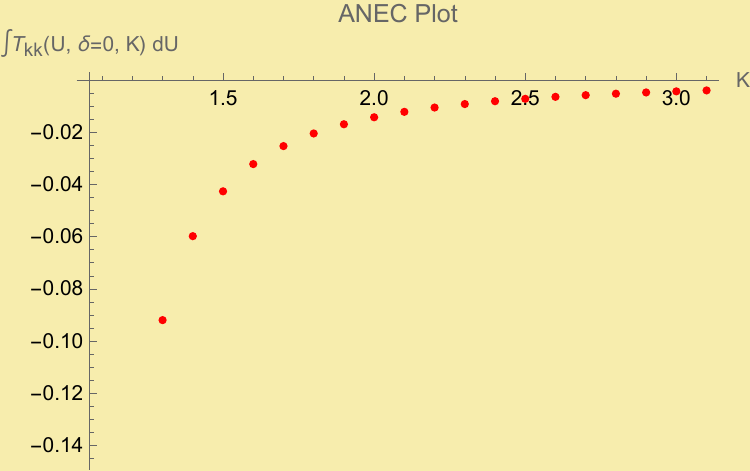}
    \end{center}
    \caption{(Left) Plot of \(T_{kk}\) versus \(U\) for various values of \(K\) when \(\delta = 0\). (Right) The integral \(\int T_{kk} dU\) plotted against \(K\).}
    \label{fig: 4D double null component for delta 0}
\end{figure}

By performing numerical integration on this full expression for massive vectors, we find that the resulting values of \(T_{kk}(U, K)\) are consistently negative, as illustrated in Figure \ref{fig: 4D double null component for delta 0}. This outcome suggests that the back-reaction helps in violation of the ANEC and makes the wormhole traversable.

In Figure \ref{fig: 4D double null component for delta 0}, the contributions to the stress-energy tensor for a massive vector field are depicted for different values of \(\phi\). A noteworthy observation is that the contribution from the massive vector field remains negative across all scenarios.

Finally, by integrating the expression for \( \langle T_{kk} \rangle \) numerically, and plotting the results against different values of \(K\), it becomes clear from Figure \ref{fig:  4D double null component for delta 0} that:
\begin{equation}
\int_0^\infty T_{kk}(U) dU < 0, \;\;\;\text{for}\;\;\;K > 1 \ .
\end{equation}

This negative value further confirms the wormhole's traversability under these conditions.

\section*{Case II: \(\delta = 2\)}

Now, consider the case where \(\delta\) is set to 2. The double null component of the stress-energy tensor in this situation is given by:

\begin{equation}
T_{kk}(U, \delta=2, K) = -\frac{45 U^2 \left(K+U^2\right)}{16 \sqrt{2} \pi  \left(K+2 U^2+1\right)^{9/2}} \, _2F_1\left(\frac{5}{2},\frac{7}{2};3;\frac{2}{2 U^2+K+1}\right)
\end{equation}

\begin{figure}[ht]
    \begin{center}
        \includegraphics[scale=0.45]{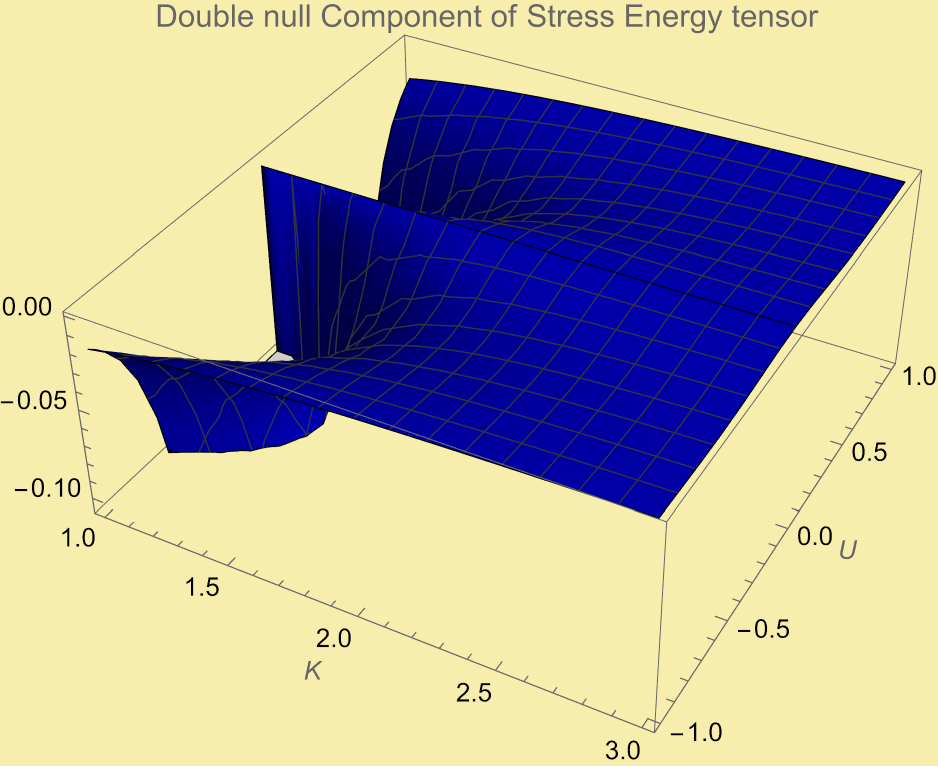}
        \hspace{0.2cm}
        \includegraphics[scale=0.72]{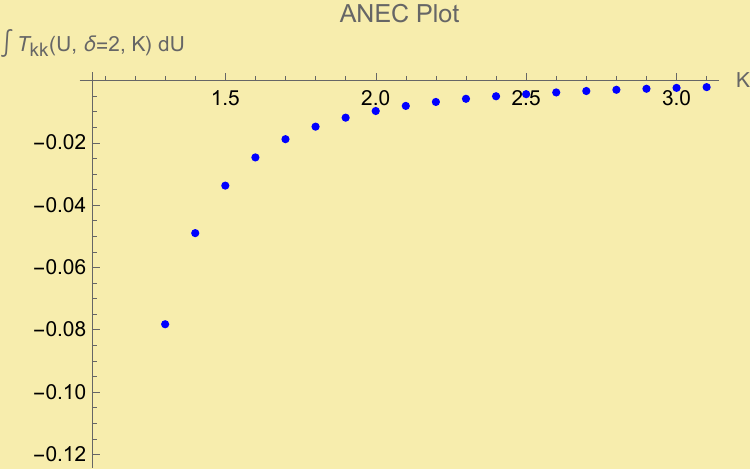}
    \end{center}
    \caption{(Left) Plot of \(T_{kk}\) versus \(U\) for various values of \(K\) when \(\delta = 2\). (Right) The integral \(\int T_{kk} dU\) plotted against \(K\).}
    \label{fig: 4D double null component for delta 2}
\end{figure}

The plot of \(T_{kk}\) as a function of \(U\) for various values of \(K\) when \(\delta = 2\) is shown in Figure \ref{fig: 4D double null component for delta 2}.

In this case, performing the ANEC integral numerically and plotting the results against different values of \(K\). From Figure \ref{fig: 4D double null component for delta 2}, it is evident that:

\begin{equation}
\int_0^\infty T_{kk}(U) dU < 0, \;\;\;\text{for}\;\;\;K > 1 \ .
\end{equation}

These results demonstrate that in the $4D$ case, the anti-periodic boundary condition leads to the traversability of the wormhole.
 \section{Wormhole in $5D$ CCBH}\label{Sec:5D CCBH wormhole}
The metric for $5D$ CCBH in Kruskal coordinate can be written as 
\begin{equation}
    ds^2 = \frac{1}{(1+UV)^2}\left[ -4 dU dV+r_h^2(-1+UV)^2 d \phi^2 + (U-V)^2 \{d\theta^2+\sin^2{\theta}d\Xi^2\}\right] \ .
\end{equation}
Using the co-ordinate defined in Appendix \ref{Appendix:t vector and parallel propagator}, it is easy to compute the geodesics distance between point $x = (U, V, \theta, \Xi, \phi)$ and $x' = (U', V', \theta', \Xi', \phi')$ :
\begin{eqnarray}
\cosh {\mu} &=&\frac{1}{(UV+1)(U'V'+1)} \Big[(UV-1)(U'V'-1)K -(U-V)(U'-V')  
    \cr && \cr
    && \;\;\;\;\;\;\;\;\;\;\;\;\;\;\;\;\;\;\;\;\;\;\;\;\; \Big\{ \sin{\theta'}\; \sin{\theta} \; \mathcal{C}_{\Xi} +\cos{\theta'} \; \cos{\theta}\Big\} +(U+V)(U'+V')\Big] \ . 
\end{eqnarray}
Here, \(K=\cosh{r_h(\phi-\phi')}\) and \(\mathcal{C}_{\Xi} = \cos(\Xi-\Xi')\).

The general expression for the two-point function as in \ref{Two point f}, and by identifying
\begin{equation*}
     \Delta_\pm = 3\pm \sqrt{1+m^2} =3 \pm \delta\;\;\;\;\;\;\;\;\;\;\;\;\;\;\;\;\;\;\;\Delta_0= \frac{7}{2} \ ,
  \end{equation*}
where $\mathcal{C}$ and $\mathcal{D}$ are
\begin{eqnarray}\label{alpha massive}
     \mathcal{C}(\mu) = \frac{1}{4}\sinh{(\mu)} \mathcal{K}'(\mu)+\cosh{(\mu)}{^{5D}\mathcal{L}(\mu)} \;\;\;;\;\;\;
     \mathcal{D}(\mu) = \frac{1}{4}\sinh{(\mu)} \mathcal{K}'(\mu)+\{\cosh{(\mu)}-1\}{^{5D}\mathcal{L}(\mu)} \ . \nonumber
 \end{eqnarray}
 the expression for ${^{5D}\mathcal{L}(\mu)}$ is
\begin{eqnarray}
    {^{5D}\mathcal{L}(\mu)} = -\frac{2^{\delta +3}  \tanh \left(\frac{\mu }{2}\right)  \left(\tanh \left(\frac{\mu }{2}\right)+1\right)^{-2 \delta } \left(\left(\delta ^2+2\right) \cosh (2 \mu )-\delta ^2+3 \delta  \sinh (2 \mu )+4\right)}{\pi ^2 e^{-5 \mu }\left(\delta ^2-1\right) \left(e^{\mu }-1\right)^6 \left(e^{\mu }+1\right)^4 (\cosh (\mu )+1)^{\delta }} \ .
\end{eqnarray}
In the case of 5D
\begin{eqnarray}
    ^{5D}\mathscr{T} (\mu, \delta) &=& \frac{\text{csch}^6\left(\frac{\mu }{2}\right) \text{sech}^4\left(\frac{\mu }{2}\right) \cosh ^2\left(\frac{\mu }{2}\right)^{-\delta } \left(\tanh \left(\frac{\mu }{2}\right)+1\right)^{-2 \delta } }{1024 \pi ^2} \Bigg[\left(34-4 \delta ^2\right) \cosh (\mu ) \nonumber \\
    && +2 \left(-2 \delta ^2+4 \cosh (2 \mu ) +\cosh (3 \mu )+8\right) + \Big\{4 \delta^2  \Big(\delta  \sinh ^3(\mu ) +\cosh (2 \mu ) \nonumber\\
    && +\cosh (3 \mu )\Big)+21 \sinh (\mu )+12 \sinh (2 \mu )+5 \sinh (3 \mu )\Big\}\Bigg]
    \cr && \cr
    ^{5D}\mathscr{G} (\mu, \delta) &=& -\frac{ \cosh ^2\left(\frac{\mu }{2}\right)^{-\delta } \left(\tanh \left(\frac{\mu }{2}\right)+1\right)^{-2 \delta } \left(\left(\delta ^2+2\right) \cosh (2 \mu )-\delta ^2+3 \delta  \sinh (2 \mu )+4\right)}{256 \pi ^2 \sinh^6\left(\frac{\mu }{2}\right) \cosh^4\left(\frac{\mu }{2}\right)}
\end{eqnarray}

Again, the expression is quite complicated, so let's look at different values of $\delta$.
 
 \quad
\section*{Case I: \(\delta = 0\)}

\begin{figure}[ht]
    \begin{center}
        \includegraphics[scale=0.45]{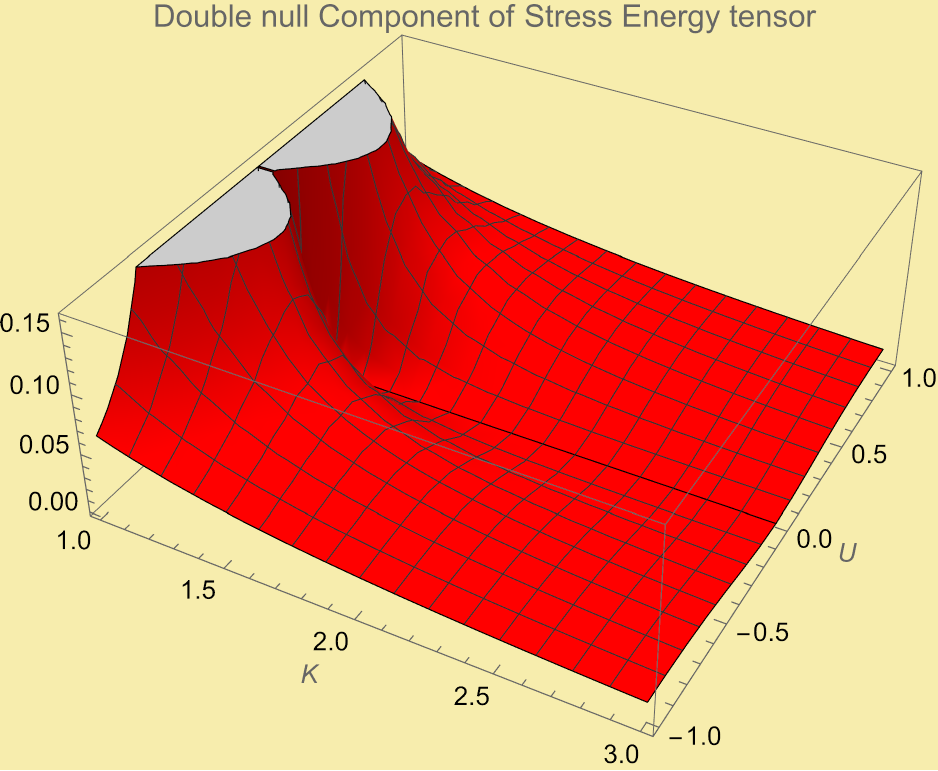}
        \hspace{0.2cm}
        \includegraphics[scale=0.69]{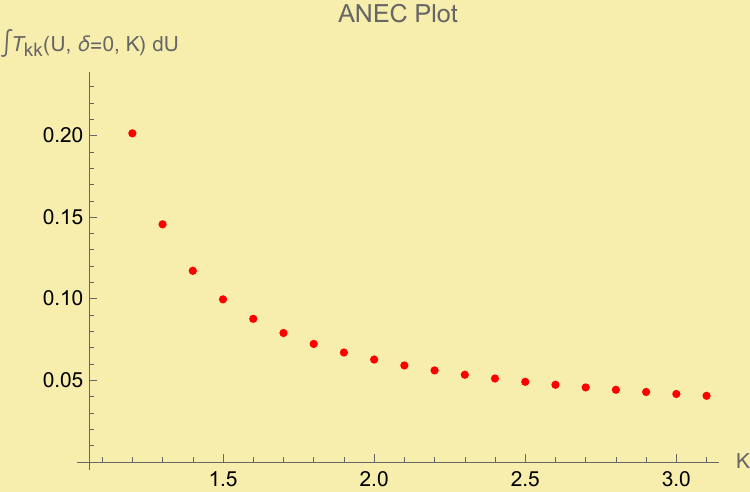}
    \end{center}
    \caption{(Left) Plot of \(T_{kk}\) as a function of \(U\) for different values of \(K\) when \(\delta = 0\). (Right) The integral \(\int T_{kk} dU\) plotted as a function of \(K\).}
    \label{fig: 5D double null component for delta 0}
\end{figure}

In this case, we consider the scenario where the parameter \(\delta\) is set to zero. Our focus is on analyzing the double null component of the stress-energy tensor. 
Upon simplification, this expression for the double null component can be written as
\[
T_{kk}(U, K) = \frac{\left(4 \mathcal{M}^3+8 \mathcal{M}^2+14 \mathcal{M}+4\right) \left(2 \mathcal{M}^2+2 \sqrt{\mathcal{M}^2-1} \mathcal{M}-\sqrt{\mathcal{M}^2-1}-\mathcal{M}-1\right) U^2}{\pi ^2 (\mathcal{M}-1)^3 (\mathcal{M}+1)^{5/2} \left(\sqrt{\mathcal{M}-1}+\sqrt{\mathcal{M}+1}\right)^3} \ ,
\]
where $\mathcal{M}=K+2U^2$. By performing numerical integration on this comprehensive expression, particularly in the context of massive vector fields, we find that the resulting values of \(T_{kk}(U, K)\) are consistently negative. This observation is depicted in Figure \ref{fig: 5D double null component for delta 0}. The negative values of the stress-energy tensor component suggest a significant physical implication: the back-reaction induced by this configuration contributes to the violation of the Averaged Null Energy Condition (ANEC). This violation, in turn, is a critical factor supporting the wormhole's traversability.

Figure \ref{fig: 5D double null component for delta 0} provides a detailed graphical representation of the contributions to the stress-energy tensor from a massive vector field, plotted for different values of \(\phi\). A particularly striking observation is that the contribution from the massive vector field remains negative across all considered scenarios, regardless of the specific values of the parameters.

To further substantiate this result, we numerically integrate the expression for \(\langle T_{kk} \rangle\) over \(U\) and plot the results against varying values of \(K\). As shown in Figure \ref{fig: 5D double null component for delta 0}, the integral satisfies the condition:
\[
\int_0^\infty T_{kk}(U) dU < 0, \quad \text{for} \quad K > 1 \ .
\]
This negative integral provides compelling evidence that the wormhole remains traversable under these specific conditions, with the violation of the ANEC playing a pivotal role.


\section*{Case II: \(\delta = 2\)}

\begin{figure}[ht]
    \begin{center}
        \includegraphics[scale=0.45]{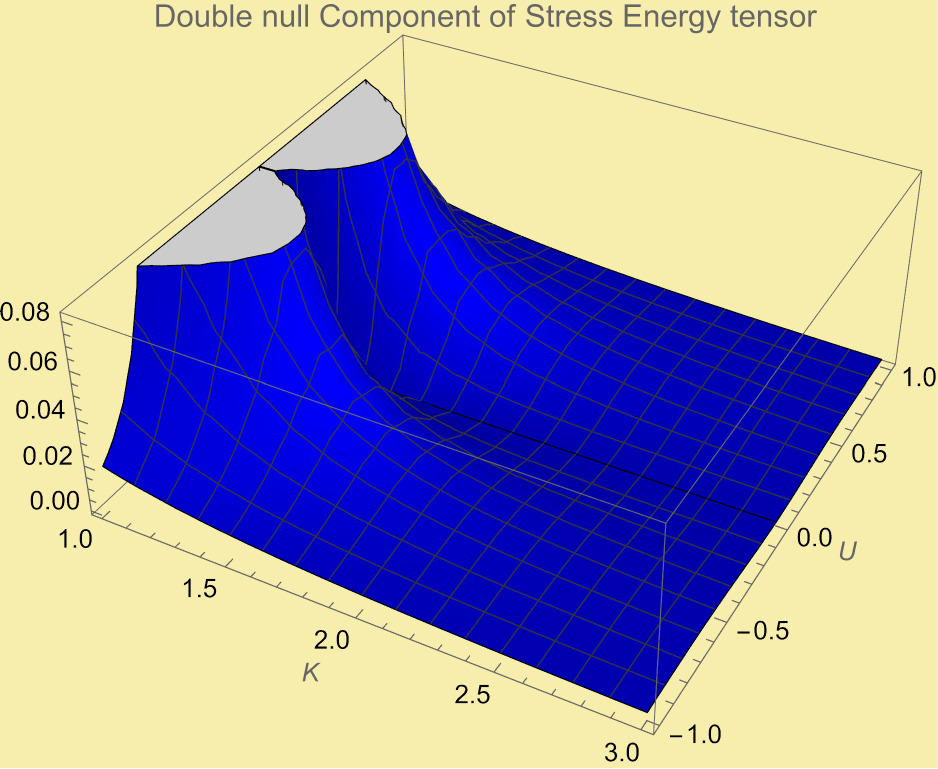}
        \hspace{0.2cm}
        \includegraphics[scale=0.72]{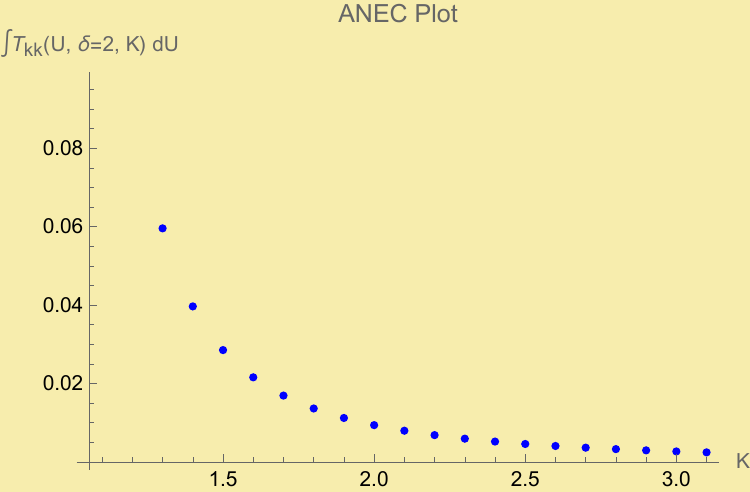}
    \end{center}
    \caption{(Left) Plot of \(T_{kk}\) as a function of \(U\) for different values of \(K\) when \(\delta = 2\). (Right) The integral \(\int T_{kk} dU\) plotted as a function of \(K\).}
    \label{fig: 5D double null component for delta 2}
\end{figure}

Now, we focus on the scenario where the parameter \(\delta\) is set to 2. In this case, the double null component of the stress-energy tensor is expressed as:
\begin{eqnarray}
T_{kk}(U, \delta=2, K) &=& \frac{24 \sqrt{\mathcal{M}+1} (3 \mathcal{M}+2)}{\pi ^2 \left(\mathcal{M}^2-1\right)^3 \left(\sqrt{\mathcal{M}-1}+\sqrt{\mathcal{M}+1}\right)^7} \Bigg(8 \mathcal{M}^4-4 \mathcal{M}^3-8 \mathcal{M}^2+3 \mathcal{M}+1 \nonumber \\
&& \;\;\;\;\;\;\;\;\;\;\;\;\;\;\;\;\;\;\;\;\;\;\;\;+\left(8 \mathcal{M}^3-4 \mathcal{M}^2-4 \mathcal{M}+1\right) \sqrt{\mathcal{M}^2-1}\Bigg) U^2 \ ,
\end{eqnarray}
where $\mathcal{M}=K+2U^2$. The behavior of \(T_{kk}\) as a function of \(U\) for various values of \(K\) when \(\delta = 2\) is illustrated in Figure \ref{fig: 5D double null component for delta 2}.

In this case, performing the numerical evaluation of the ANEC integral and plotting the results against different values of \(K\) provides us with crucial insights. As seen in Figure \ref{fig: 5D double null component for delta 2}, the integral over the double null component satisfies the following condition:

\[
\int_0^\infty T_{kk}(U) dU < 0, \quad \text{for} \quad K > 1 \ .
\]

These findings confirm that, in the five-dimensional (5D) context, applying an periodic boundary condition leads to the violation of the ANEC. This violation is a key factor that allows the wormhole to be traversable, highlighting the significance of the underlying physical mechanisms in enabling such exotic spacetime structures.

\section{ Conclusion}\label{Sec:Conclusion}

This work explores the feasibility of traversable wormholes in the context of a particular \(\mathbb{Z}_2\) symmetry quotient of Constant Curvature Black Hole (CCBH) spacetimes sustained by a substantial gauge field. We initiated a comprehensive examination of the CCBHs, laying the foundation for our analysis. We computed the geodesic distances in $AdS$ spacetimes to analyze the fundamental geometric structure. As results, the calculations were essential for ascertaining the tangent vectors and the parallel propagators, and with their help, we also computed the expression for the two-point function related to massive gauge fields. Subsequently, 
the double null component of the stress-energy tensor and assessed the Averaged Null Energy Condition (ANEC) are calculated, thereby establishing that ANEC is violated within this framework. The breach of ANEC leads to a notable outcome: the backreaction on the spacetime geometry caused by our configuration enhances the traversability of the wormholes. Given the examined parameters, this observation highlights the possible presence of traversable wormholes in the present work.

Furthermore, our findings provide essential corroboration of earlier assertions. The authors of \cite{Fu:2018oaq} proposed that a particular selection of boundary conditions, i.e., periodic and anti-periodic in their framework, might facilitate the violation of the ANEC principle. In contrast, our study demonstrates that the selection is not random, i.e., the choice of the relative sign in Eq.\eqref{State relation} is not arbitrary. The outcome is contingent upon the dimensions of spacetime under consideration and the spin characteristics of the particles involved. This selection adheres to the relationship $(-1)^{Ds}$, where \(D\) signifies the dimensionality of spacetime and \(s\) indicates the spin of the quantum fields. As demonstrated in \cite{Anand:2020wlk}, implementing anti-periodic boundary conditions violates the Averaged Null Energy Condition (ANEC). For our general selection, we have considered the cases of different spacetime dimensions in this study. Moreover, it was noted that for \(\delta = 1\), the formulation of the double null component exhibits divergence, thereby uncovering complex dynamics of the stress-energy tensor in particular scenarios. This divergence underscores the intricate nature of the underlying principles, illustrating its reliance on selected parameters and setting the stage for additional explorations into the characteristics of traversable wormholes. The extension of this investigation to examine the potential traversability of wormholes in the context of Gravitino and Graviton fields presents a compelling area of inquiry. Furthermore, analyzing traversability in the context of fermions warrants careful consideration. Recent studies indicate that Euclidean wormholes offer significant perspectives on the information loss paradox. Consequently, it would be fascinating to ascertain if the results concerning traversability in Lorentzian wormholes, as investigated in this study, offer any significant insights into this specific problem. We intend to re-examine this subject in subsequent investigations.

\subsection*{Acknowledgments}

AA would like to express his deepest gratitude to Prasanta Tripathy for their invaluable guidance, constant support, and insightful feedback. He is truly fortunate to have had the opportunity to learn a lot from him. DS is supported by the Fundamental Fund of Khon Kaen University. DS has also received funding support from the National Science, Research and Innovation Fund and supported by Thailand NSRF via PMU-B [grant number B37G660013]. 




\begin{appendices}


\section{The co-ordinate choice for 4D and 5D CCBH black holes}\label{Appendix:t vector and parallel propagator}
\subsection*{The CCBH in $4D$ Case}

The choice of co-ordinate for 
\begin{eqnarray} \label{Embedding}
T_1 &=& \frac{r}{r_h}\cosh{\left(r_h \phi\right)} =  \frac{1-UV}{1+UV} \cosh{\left( r_h \phi \right)} 
\cr && \cr
T_2 &=&\frac{\sqrt{r^2-r_h^2}}{r_h} \sinh{\left(r_h t\right)}  =  \frac{U+V}{1+UV}
\cr && \cr
X_1 &=& \frac{r}{r_h}\sinh{\left(r_h \phi\right)} = \frac{1-UV}{1+UV} \sinh{\left( r_h \phi \right)}  
\cr && \cr
X_2 &=& \frac{\sqrt{r^2-r_h^2}}{r_h} \cosh{\left(r_h t \right)} \cos{\theta} = \frac{V-U}{1+UV} \cos{\theta} 
\cr && \cr
X_3 &=& \frac{\sqrt{r^2-r_h^2}}{r_h} \cosh{\left(r_h t\right)} \sin{\theta}  =  \frac{V-U}{1+UV} \sin{\theta} \ .
\end{eqnarray} 
The metric in the Schwarzschild coordinates $(t,r,\theta,\Xi,\phi)$ is 
\begin{equation}
    ds^2 = -(r^2-r_h^2)dt^2+\frac{dr^2}{r^2-r_h^2}+\frac{r^2-r_h^2}{r_h^2}\cosh^2{r_h t} d\theta^2  + r^2 d\phi^2 \ .
\end{equation}
Here, \( r = r_h \) denotes the location of the black hole horizon.


\subsection*{The CCBH in $5D$ Case }
Similar to the $4D$ case, one can have 
\begin{eqnarray} \label{Embedding5D}
T_1 &=& \frac{r}{r_h} \; \cosh{r_h \phi} =  \frac{1-UV}{1+UV} \;  \cosh{ r_h \phi } 
\cr && \cr
T_2 &=&\frac{\sqrt{r^2-r_h^2}}{r_h} \; \sinh{r_h t}  =  \frac{U+V}{1+UV}
\cr && \cr
X_1 &=& \frac{r}{r_h} \; \sinh{r_h \phi} = \frac{1-UV}{1+UV} \;  \sinh{ r_h \phi }  
\cr && \cr
X_2 &=& \frac{\sqrt{r^2-r_h^2}}{r_h} \; \cosh{ r_h t } \; \cos{ \theta} = \frac{V-U}{1+UV} \; \cos{ \theta } 
\cr && \cr
X_3 &=& \frac{\sqrt{r^2-r_h^2}}{r_h} \; \cosh{r_h t}  \; \sin{\theta } \; \cos{\Xi}  =  \frac{V-U}{1+UV} \; \sin{ \theta} \; \cos{\Xi} \ ,
\cr && \cr
X_4 &=& \frac{\sqrt{r^2-r_h^2}}{r_h} \; \cosh{r_h t}  \; \sin{\theta }  \; \sin{\Xi}  =  \frac{V-U}{1+UV}  \; \sin{ \theta} \; \sin{\Xi} \ .
\end{eqnarray}
The metric in the Schwarzschild coordinates $(t,r,\theta,\Xi,\phi)$ is 
\begin{equation}
    ds^2 = -(r^2-r_h^2)dt^2+\frac{dr^2}{r^2-r_h^2}+\frac{r^2-r_h^2}{r_h^2}\cosh^2{r_h t}\left(d\theta^2 + \sin^2{\theta}\; d\Xi^2 \right) + r^2 d\phi^2 \ .
\end{equation}
Here, \( r = r_h \) denotes the location of the black hole horizon.

\end{appendices}




\end{document}